\documentclass[10pt,paper]{revtex4}
\usepackage{graphicx}

\begin{document}

\title{Mass redistribution in variable mass systems}
\author{C\'elia A. de Sousa and V\'\i  tor H. Rodrigues}
\affiliation{Departamento de F\'\i  sica da Universidade de Coimbra,\\ P-3004-516 Coimbra, Portugal}
\email{celia@teor.fis.uc.pt}

\begin{abstract}
We have developed an alternative formulation based on ${\bf F}\,=\,M\,{\bf a}$ rather than  ${\bf F}\,=\,d{\bf
P}/dt$  for studying   variable mass systems. It is shown that ${\bf F}\,=\,M\,{\bf a}$ can be particularly useful
in this context, as illustrated by various examples involving chains and ropes. The method implies the division of
the whole system into two parts, which are considered separately, allowing to explore certain aspects as
constraint forces.
\end{abstract}
\maketitle
 \section {Introduction }


Articles on teaching of variable mass systems have addressed a number of interesting  issues \cite{Thorpe, AronsB,
Tiersten, Siegel, Krane}. For instance, through adequate  examples, Tiersten \cite {Tiersten} illustrates a
sophisticated approach to be applied to open systems,  i.e., systems for which there exists a mass influx or
efflux. However, in some cases many questions still remain, such as the definition of the {\em system} and a
clarification of the terminology which is  used.  In addition, the tendency to assume that ${\bf F}\,=\,d\,{\bf
P}/dt$ is, contrary  to ${\bf F}\,=\,M\,{\bf a}$, always applicable to a variable mass system is well documented
and generalized among students. In our opinion these facts are strongly correlated to the way this subject is
taught in  introductory courses.

In this paper we will suggest a different approach, where    ${\bf F}\,=\,M\,{\bf a}$ plays  a central role. As
the whole (closed) system under study is separable  in two (variable mass) subsystems, we will show that a general
equation of the type ${\bf F}\,=\,M\,{\bf a}$ can be used  as  the equation of motion for  one or, in certain
cases, for both parts of the system.

 During the motion, mass is  interchanged between    both systems although  the
mass of the total system remains constant. Hence, mass is exchanged between the two subsystems and this has to be
kept in mind in writing down the  equations of motion. One may choose the {\em system} arbitrarily, including or
excluding part of the total number of particles, but if interactions act through the system boundary, this must be
taken into account. Understanding such distinctions between different choices of system plays an important role in
the subsequent discussion.

In this context, we develop  a method  to study variable mass systems which is more clear than the elementary
traditional method \cite{Benson}. Three sample problems concerning nonrigid systems (ropes and chains) are
presented in detail to illustrate  the pedagogical value of the method.

We have verified in the classroom that these problems, directly usable in courses of mechanics,  provide useful
insights into the applicability of multi-particle forms of Newton's second law, allowing the teacher to revisit
key concepts of dynamics.

The great variety of systems that can be studied by the proposed methodology, such as ropes, chains or conveyor
belts, shows that it can be incorporated in the background required in the scientific formation of experts of
engineering activities. Students should be constantly alert to the various assumptions and approximations in the
formulation and solution of real problems. Bearing in mind that certain approximations will always be necessary,
the ability to understand the physical context  and to construct   the idealized mathematical model for some
engineering problems is a crucial formative aspect. For instance, the weight of a cable (or a rope) per unit
length may be neglected if the tension in the cable  is much greater than its total weight, whereas the cable
weight may not be neglected so far as the calculation of the deflection of a suspended cable due to its weight is
concerned. The formulation and analysis of practical problems involving the principles of  dynamics can give the
bases  to develop systematically these abilities.
 \section {General Equations of Motion }
Newton's second law states that the rate of change of the momentum  ${\bf P}$ of a  {\em closed}  system of mass
$M$  is equal to the net external force ${\bf F}$ acting upon it:

\begin{equation}\label{eq:dp}
{\bf F}\,=\,\frac{{\rm d}{\bf P}}{{\rm d}\,t}.
\end{equation}

Alternatively,  use can be made of

\begin{equation}\label{eq:ma}
{\bf F}\,=\,M\,{\bf a}_{\rm cm},
\end{equation}
where   ${\bf a}_{\rm cm}$ is the acceleration of the centre of mass of the system.

If the problem involves variable mass systems, authors  assume that it is more convenient to work with momentum,
and start by using  (\ref {eq:dp}). Of course, under the same conditions, the second statement (\ref {eq:ma}) can
also be used.

\subsection {Standard elementary approach}
The traditional method \cite{Benson} to obtain the equation of motion of a variable   mass system makes use of
(\ref{eq:dp}). The system is supposed to have  a principal part of mass $M$ moving with velocity ${\bf v}$, and a
body of mass $\Delta \,M$ moving with velocity ${\bf  u}$ $({u}>{v}$), which undergo a completely inelastic
collision. After the collision the body of mass $M\,+\,\Delta\,M$ is assumed to move  with a  velocity ${\bf
v}\,+\,\Delta\,{\bf v}$, so that the variation of   linear momentum is $(M+\Delta M)\,({\bf v}+ \Delta {\bf
v})\,-\,(M {\bf v}\,+\,\Delta M {\bf u})$. Equating the increase in momentum and  the impulse of the external
force ${\bf F}$ acting upon   the {\em total} (constant mass) system, and dividing both sides of the equation  by
$\Delta t$, we obtain, in the limit $\Delta t \rightarrow 0$,
\begin{equation}\label{eq:ben}
\frac{{\rm d}}{{\rm d}\,t}\,(M\,{\bf v})\,=\,{\bf F}\,+\,{\bf u}\,\frac{{\rm d}\,M}{{\rm d}\,t}.
\end{equation}

Here,  ${\rm d}M/{\rm d}t$ is the rate at which the mass enters in the principal part of the system.

\subsection {Proposed approach and its advantages}
Let us consider a main body which undergoes incremental mass change, increasing (or decreasing) its mass. The mass
of this  body (subsystem II) and its velocity at an arbitrary instant $t$ are $M$ and $ {\bf v}$, respectively, as
it is shown in figure 1. The rest of the system (subsystem I) has mass $m$, moving with velocity ${\bf u}$
(${u}>{v}$) at the same instant. So,  as the motion progresses the two subsystems I and II coalesce to one
another. This means that the body I of mass $m$ loses mass which immediately becomes part of the main body II. We
should keep in mind that $m$ and $M$ are time dependent, but the whole system has  {\em constant} total mass
$m+M$, so that mass is being transferred at the rate ${\rm d}\,M/{\rm d} t\,=\,-\,{\rm d}\,m/{\rm d} t$.

\begin{figure}[h]
\begin{center}
\scalebox{0.60}{\includegraphics{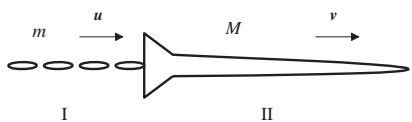}}
\caption{\label{fig:sousafig1} Schematic diagram for the two
subsystems I and II.}
\end{center}
\end{figure}
  Newton's second law   for the total {\em closed system} reads
\begin{equation}\label{eq:map}
{\bf {\cal F}}\,=\,( m\,+ M)\,{\bf a}_{\rm cm},
\end{equation}
if we adopt the point of view of  (\ref{eq:ma}).  ${\cal F}$ is the external force acting upon  the total system.

 The centre of mass velocity is
\begin{equation}
{\bf v}_{\rm cm}\,=\,\frac{1}{m+M}\,\left(m\,{\bf u}\,+\,M\,{\bf v}\right).
\end{equation}

The time derivative of this expression  ($m+M=$ constant, ${\rm d}m/{\rm d}t=-{\rm d}M/{\rm d}t$) is  the centre
of mass acceleration:

\begin{equation}
{\bf a}_{\rm cm}\,=\,\frac{1}{m+M}\,\left(m\,\frac{{\rm d}{\bf u}}{{\rm d}t}\,-\,\frac{{\rm d}M}{{\rm d}t}\,{\bf
u}\,+\,M\, \frac{{\rm d}{\bf v}}{{\rm d}t}\,+\,\frac{{\rm d}M}{{\rm d}t}\,{\bf v}\,\right).
\end{equation}

The insertion of this expression in (\ref{eq:map}) led us directly to

\begin{equation}\label{eq:tot}
{\bf {\cal F}}\,=\,{\bf F}^{({\rm I})}\,+\,{\bf F}^{({\rm II})}\,=\,m\,\frac{{\rm d}{\bf u}}{{\rm
d}t}\,-\,\frac{{\rm d}M}{{\rm d}t}\,{\bf u}\, +\,M\,\frac{{\rm d}{\bf v}}{{\rm d}t}\,+\,\frac{{\rm d}M}{{\rm
d}t}\,{\bf v}\,,
\end{equation}
where ${\bf F}^{({\rm I})}$ and ${\bf F}^{({\rm II})}$ are the  forces acting upon subsystems I and II,
respectively.

In  order to describe  separately the  motion of each subsystem I and II, we split  (\ref{eq:tot}) into two:

\begin{equation}\label{eq:mav1}
{\bf F}^{({\rm I})}\,=\,m\,\frac{{\rm d}{\bf u}}{{\rm d}t},\hskip0.2cm {\rm for\, the\, variable\, mass\, system
\,I,}\,
\end{equation}
and
\begin{equation}\label{eq:mav2}
{\bf F}^{({\rm II})}\,=\,M\,\frac{{\rm d}\,{\bf v}}{{\rm d}\,t}\,+\,({\bf v}\,-\,{\bf u})\,\frac{{\rm d}\,M}{{\rm
d}\,t},\hskip0.2cm {\rm for\, the\, variable\, mass\, system \,II}\,.
\end{equation}

This splitting of  (\ref{eq:tot}) into (\ref{eq:mav1}) and (\ref{eq:mav2}) is physically fruitful as is
illustrated in section III.

The role of the systems I and II can be interchanged, and both of them should be described by the standard
equation for a variable mass system. With this goal in mind, it is convenient to rewrite    (\ref{eq:mav1}) and
(\ref{eq:mav2})  using the linear momentum.

From  (\ref{eq:mav1}) we obtain
\begin{equation}\label{eq:vari}
{\bf F}^{({\rm I})}\,=\,\frac{{\rm d}}{{\rm d}t}\,(m\,{\bf u})\,-\,{\bf u}\,\frac{{\rm d}\,m}{{\rm d}t},
\end{equation}
and, from  (\ref{eq:mav2}),
\begin{equation}\label{eq:varii}
{\bf F}^{({\rm II})}\,=\,\frac{{\rm d}}{{\rm d}t}\,(M\,{\bf v})\,-\,{\bf u}\,\frac{{\rm d}\,M}{{\rm d}t},
\end{equation}
proving the  coherence of the treatment. Both expressions have a common generic structure hence, hereafter,  the
symbols I or II are dropped.

 In conclusion, the standard Newton's  equation of motion for variable mass systems  is obtained:
\begin{equation}\label{eq:var}
\frac{{\rm d}\,{\bf P}}{{\rm d}\,t}\,=\,{\bf F}\,+\,{\bf u}\,\frac{{\rm d}\,M}{{\rm d}\,t}.
\end{equation}

It is worth  pointing out that:
\begin{itemize}
\item  the {\em system} has instantaneous mass $M$ ($m$), and  linear momentum ${\bf
P}\,=\,M\,{\bf v}$ ($m\,{\bf u}$);
\item   ${\bf F}$   is the net external force acting upon the {\em variable} mass system;
\item ${\bf u}\,{{\rm d}M}/{{\rm d} t}$ (${\bf u}\,{{\rm d} m}/{{\rm d} t}$) is the rate at which momentum is
carried into or away from the system of mass $M$ ($m$).
\end{itemize}

In the present derivation, the definition of the {\em system} is clear, as well as the concept of linear momentum
which agrees with the concept students are acquainted with. Regarding the approach presented previously, we notice
that ${\bf F}$ (see (\ref{eq:ben})) is the force acting on the whole (constant mass) system.

 This approach is suitable for undergraduate students  and is equivalent to other
 methods \cite{Tiersten,MaMa}.

\section{Examples}

There are many one-dimensional nonrigid systems, such as chains and ropes, for which the usage of conservation
laws of energy and linear momentum represent a good approach. However, the conservation laws  {\em always} refer
to a definite number of particles, although the system under study might be split  into two subsystems of variable
mass. It is this   aspect that is central in  our approach. In this context, three illustrative examples are
discussed.

\subsection { Example 1: Falling of a chain}

 {\em The upper end of a uniform  open-link chain of length $l$, and  mass per unit length
$\lambda$, is released from rest at $x=0$. The lower end is fixed at point A as it is shown in figure 2. Find the
tension $T(x)$  in the chain at point A after the upper end of the chain has dropped the distance  $x$. Assume the
free fall of the chain.}

\begin{figure}[th]
\begin{center}
\scalebox{0.70}{\includegraphics{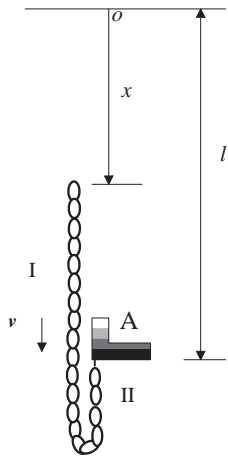}}
 \caption{The chain with the displacement $x$ of the top end.}
\end{center}
\end{figure}
\vskip0.2cm
 {\bf Solution.} According with the suggested methodology, we   divide the chain into two parts each
to be treated
 using Newton's second law for variable mass systems  (\ref{eq:var}).
For this purpose we consider:
\begin{itemize}
\item {\bf subsystem I}: the free falling  part of the chain, whose mass is $\lambda (l-{x}/{2})$, and   velocity $v$;
\item {\bf subsystem II}: the rest of the chain, whose   mass is $\lambda\,x/2$, and at rest.
\end{itemize}

Accordingly, for {\bf subsystem I}:

\begin{equation}
M\,=\,\lambda\,(l\,-\,\frac{x}{2}),\hskip0.5cm P\,=\,\,\lambda\,(l\,-\,\frac{x}{2})\,v,\hskip0.5cm {\rm and}
\hskip0.5cm u=v.
\end{equation}

\begin{table}[h]
\begin{center}
\begin{tabular}{|c|c|c|c|c|c|}
\hline
System& $M$&$P$&$F$&$u$& $u\,{\rm d}M/{\rm d}t$ \\
\hline \hline
I &$\lambda\,(l\,-\,{x}/{2})$ &$\lambda\,(l\,-\,{x}/{2})\,v$& $\lambda\,(l\,-\,{x}/{2})\,g$& $v$ &$-\lambda\,v^2/2$\\
\hline
II &$\lambda\,\,x/2$ &$0$& $\lambda\,x\,g/2+\,T\,(x)$&  $v$  &$\,\lambda\,v^2/2$\\
\hline
I+II &$\lambda\,l$ &$\lambda\,(l\,-\,{x}/{2})\,v$& $\lambda\,l\,g+\,T(x)$&{---}&{---}\\
\hline
\end{tabular}
\caption{\label{tab:(1)} Summary of some relevant physical quantities for example 1. See  text for notation.}
\end{center}
\end{table}

These quantities are displayed in table I, as well as the corresponding quantities for subsystem II and the total
system.

 As the net external force is
 $F\,=\,\lambda\,(l\,-\,x/2)\,g$, where $g$ stands for the gravity acceleration, equation (\ref{eq:var})  can be
 written as

\begin{equation}
\frac{{\rm d}}{{\rm d}t}\,[\lambda\,(l\,-\,\frac{x}{2})\,v]\,=\,\lambda\,(l\,-\,\frac{x}{2})\,g\,-\,
\frac{\lambda}{2}\,v^2.
\end{equation}

After straightforward  simplifications one  concludes that $a\,=\,{\rm d}v/{\rm d}t\,=\,g$,  which is consistent
with the free fall assumption already used to write $F$.

The velocity can then be  calculated from  ${\rm d}v/{\rm d}t\,=v \,{\rm d}v/{\rm d}x\,=\,g$. The  first-order
differential equation  $v\,{\rm d}v\,-\,g\,{\rm d}x=0$ is  integrated ($x_0=v_0=0$) yielding
\begin{equation}\label{eq:vx}
v\,(x)\,=\,(2\, g\, x)^{1/2}.
\end{equation}

As stated above, one can also confirm that  a general equation of the type ${\bf F}\,=\,M\,{\bf a}$ applies to
subsystem I ($F=\lambda\,(l\,-\,x/2)\,g$, $M=\lambda\,(l\,-\,x/2)$, $a=g$). This is a consequence of having ${
v}\,=\,{ u}$. In this case mass is being {\em lost} but at zero relative velocity.

 To proceed, we apply  the reasonings used above to {\bf subsystem II}.   As    the net external force on this subsystem is
 $F\,=\,T\,(x)\,+\,\lambda\,x\,g/2$,  one  finds 
(see table I), from  (\ref{eq:var})

 \begin{equation}
 0\,=\,T\,(x)\,+\,\frac{1}{2}\lambda\,x\,g\,+\,\lambda\,\frac{v^2}{2}.
 \end{equation}

 Inserting (\ref{eq:vx}) we obtain for the tension
\begin{equation}\label{eq:tx1}
T\,(x)\,=\,-\,\frac{3}{2}\,\lambda\,x\,g.
\end{equation}

One may now check  this result using  (\ref{eq:ma}) for the whole  {\bf system I+II}, and to this end the
acceleration of the centre of mass  must be obtained.

 From figure 2 and table I one has for the centre of mass velocity

\begin{equation}\label{eq:vcm}
v_{\rm cm}\,=\,v\,(1\,-\,\frac{x}{2 l}).
\end{equation}

Taking the  time derivative we get the  centre of mass acceleration:
\begin{equation}
a_{\rm cm} \,=\,-\,\frac{v^2}{2l}\,+\,g\,(1\,-\,\frac{x}{2 l}).
\end{equation}

Inserting now $v(x)$ as given by   (\ref{eq:vx}) in this  equation yields
\begin{equation}\label{eq:acm}
a_{\rm cm}\,=\,g\,(1\,-\,\frac{3 x}{2l}).
\end{equation}

Newton's second law (\ref{eq:ma}), applied to the chain   (closed system I+II),  reads (see table I)

\begin{equation}
\lambda\,l\,g\,+\,T\,(x)\,=\,\lambda\,l\,g\,(1\,-\,\frac{3 x}{2 l}),
\end{equation}
 which confirms  (\ref{eq:tx1}) for $T\,(x)$.

This kind of nonrigid systems is also accurate to study energetic processes \cite{Sousa}. In fact, the initial
mechanical energy of the chain is converted into   internal energy.

\subsection {Example 2: Falling of a rope}

{\em A coil of a uniform rope is placed just above a hole in a platform. One end of the rope falls (without
friction or air resistance)  and pulls down the remaining rope   in a steady motion (see figure 3). The rope has
total length $l$ and mass per unit length $\lambda$, and  starts from rest at $x=0$. Find the normal force $N(x)$
exerted upon the coil, and the tension $T(x)$ in the rope at a distance $x$ from its lower end. The system is
confined in a box so that the rope is prevented from lifting up.}
\begin{figure}[h]
\begin{center}
\scalebox{0.60}{\includegraphics{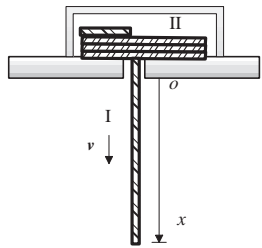}}
\caption{\label{fig:sousafig3} Scheme of the rope falling off
with $x$ being the length of the rope  hanging from the hole.}
\end{center}
\end{figure}

\vskip0.2cm
 {\bf Solution.} The rope behaves like a flexible and inextensible one-dimensional conservative system.
Under these conditions, as it falls down the mechanical energy is conserved. The acceleration of the falling rope
is not known, as well as the   forces $N(x)$ and $T(x)$. Notice that these forces, which  result from the
constraints imposed on the motion of the rope, cannot be specified {\em a priori}.

The acceleration of the falling rope may be obtained form the {\em energy conservation}.
\begin{table}[h]
\begin{center}
\begin{tabular}{|c|c|c|c|c|c|}
\hline
System& $M$&$P$&$F$&$u$& $u\,{\rm d}M/{\rm d}t$ \\
\hline \hline
I &$\lambda\,x$ &$\lambda\,x\,v$& $\lambda\,x\,g+\,T(x)$& $v$ &$\lambda\,v^2$\\
\hline
II &$\lambda\,(l\,-\,x)$ &$0$& $\lambda\,(l\,-\,x)\,g+\,N\,(x)\,-\,T(x)$&  $v$  &$-\,\lambda\,v^2$\\
\hline
I+II &$\lambda\,l$ &$\lambda\,x\,v$& $\lambda\,l\,g+\,N(x)$&{---}&{---}\\
\hline
\end{tabular}
\caption{\label{tab:(2)} Summary of some relevant physical quantities for example 2. See  text for notation.}
\end{center}
\end{table}

As illustrated in figure 3, the $x$ axis, whose   origin is on the platform, points downwards.  The variation of
the kinetic energy of the system is

\begin{equation}
\Delta\,K\,(x)\,=\,\frac{1}{2}\,\lambda\,x\,v^2,
\end{equation}
where $v={\rm d}x/{\rm d}t$ is the velocity of the moving part of the rope.

The corresponding variation of the potential energy (whose zero is for $x=0$) is
\begin{equation}
\Delta U(x)\,=\,-\,\frac{1}{2}\,\lambda\,g\,x^2.
\end{equation}

From the conservation of energy, $\Delta\,E=0$,  we obtain  the velocity, $v$, as a function of    $x$,
\begin{equation}\label{eq:vx2}
v(x)\,=\,(g\,x)^{1/2}.
\end{equation}

Taking the time derivative of this equation the acceleration of the falling rope is obtained:
\begin{equation}\label{eq:ace2}
a\,=\,\frac{{\rm d}v}{{\rm d}t}\,=\,\frac{{\rm d}v}{{\rm d}x}\,\frac{{\rm d}x}{{\rm d}t}\,=\,\frac{g}{2}.
\end{equation}

We now illustrate the recommended approach, by looking at the problem from the point of view of a variable mass
system.

 As figure 3 shows  the system at time $t$  can be separated into two subsystems:
\begin{itemize}
\item {\bf subsystem I}: the hanging  part  with length $x$,  mass $\lambda\,x$ and velocity $v$;
\item {\bf subsystem II}:  the remainder rope  at rest on the platform,  with mass $\lambda\,(l-x)$.
\end{itemize}

Since we are interested to study  the rope as a variable mass  system, we put it inside  a box, as shown in figure
3. Therefore the rope will not depart from the assumed shape as soon as the motion starts.

For  {\bf subsystem I} (see table II, and figure 3) one has:

\begin{equation}
M=\lambda \,x,\hskip.5cm P=\lambda\,x\,v,  \hskip0.5cm u=v,\hskip0.5cm {\rm and} \hskip0.5cm u\,\frac{{\rm
d}M}{{\rm d}t}\,=\,\lambda\,v^2.
\end{equation}

Because the  external force acting on this system is $F=\lambda\,x\,g\,+\,T(x)$, where $T(x)$ is the tension in
the rope at the boundary of  subsystem I,  (\ref{eq:var}) leads to
\begin{equation}
\frac{{\rm d}}{{\rm d}t}\,(\lambda\,x\,v)\,=\,\lambda\,x\,g\,+\,T(x)\,+\,\lambda\,v^2,
 \end{equation}
and, after  using  (\ref{eq:ace2}),

\begin{equation}\label{eq:tx2}
T\,(x)\,=\,\lambda\,x\,(a\,-\,g)\,=\,-\,\frac{1}{2}\lambda\,x\,g.
\end{equation}

 As we have already said,    a general equation
of the type ${\bf F}\,=\,M\,{\bf a}$ is also valid for this part of the system ($F\,=\,\lambda\,x\,g/2$,
$M=\lambda\,x$, $a={\rm d}v/{\rm d}t\,=\,g/2$). As in example 1, also here the mass is {\em acquired} at zero
relative velocity.

A similar study for {\bf subsystem II} can be carried out. The relevant physical quantities for this variable mass
system are also given in table II.

 As the net external
force on subsystem II  includes the  weight of part of the rope, $\lambda\,(l\,-\,x)\,g$, the tension,
$-\,T\,(x)\,=\,\lambda\,x\,g/2$, and the total force by the platform, $N(x)$, the equation of motion
(\ref{eq:var}) yields
\begin{equation}
0\,=\,\lambda\,(l\,-\,x)\,g\,+\frac{1}{2}\,\lambda\,x\,g\,+\,\,N\,(x)\,-\,\lambda\,v^2.
\end{equation}

Using    (\ref{eq:vx2}) the normal force on the rope can be obtained:
\begin{equation}\label{eq:nx2}
N\,(x)\,=-\lambda\,(l\,-\,\frac{3}{2}\,x)\,g,
\end{equation}
showing  that $N(x)=0$ when $x=2l/3$. We may now confirm  this result by applying Newton's second law to the
closed {\bf system I+II}  using  (\ref{eq:ma}).

The  velocity of the centre of mass can be calculated directly from the quantities mentioned  in figure 3 and in
table II:
\begin{equation}\label{eq:vcm1}
v_{\rm cm}\,=\,\frac{x}{l}\,v\,=\,\frac{x}{l}\,(g\,x)^{1/2}.
\end{equation}

The time derivative of this equation, together with  (\ref{eq:vx2}) and (\ref{eq:ace2}), allows us calculate the
acceleration of the centre of mass
\begin{equation}\label{eq:acm2}
a_{\rm cm}\,=\,\frac{{\rm d}\,v_{\rm cm}}{{\rm
d}\,t}\,=\,\frac{v^2}{l}\,+\,\frac{x}{l}\,a\,=\,g\,\frac{3\,x}{2\,l}.
\end{equation}

This equation   shows that $a_{\rm cm}$ becomes equal to $g$ when $x=2\,l/3$, and agrees with $N\,(x)\,=\,0$  for
that value of $x$.

Newton's second law (\ref{eq:ma}) applied to  the  { system I+II} (see  table II) leads to
\begin{equation}
\lambda\,l\,g\,+\,N\,(x)\,=\,\lambda\,l\,(\,\frac{3\,x}{2\,l}\,g),
\end{equation}
resulting a  total force   on the rope as given by  (\ref {eq:nx2}), showing  the validity of the suggested
method.

\subsection {\bf Example 3: Pulling a rope}

{\em A uniform rope, of length $l$ and mass per unit length $\lambda$, is pulled along a smooth horizontal surface
by a constant force $F$. Find the tension in the rope at any point a distance $x$ from the end where  $F$ is
applied.}
\begin{figure}[h]
\begin{center}
\scalebox{.50}{\includegraphics {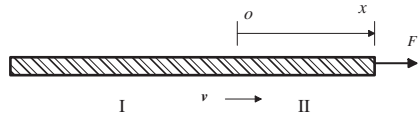}} \caption{\label{fig:sousafig4} Configuration of the rope being
pulled on a smooth horizontal surface by the constant force $F$.}
\end{center}
\end{figure}
\vskip0.2cm

{\bf Solution.}  Using again the variable mass problem-solving perspective, we present in table III (see also
figure 4) some of the relevant physical quantities for this example. The origin of the reference system is such
that $x$ is the coordinate of  the point of application of $F$.

The equation of motion (\ref{eq:var}), applied to {\bf subsystem I}, is written as
\begin{equation}\label{eq:sii}
\frac{{\rm d}}{{\rm d}t}\,[\lambda\,(l\,-\,x)\,v]\,=\,T\,(x)\,-\,\lambda\,v^2,
\end{equation}
where $T(x)$ is the tension in the rope at a distance $x$ from the front top of the rope. The  terms involving the
velocity cancel in both sides of  (\ref{eq:sii}) and therefore
\begin{equation}\label{eq:tx3}
T\,(x)\,=\,\lambda\,(l\,-\,x)\,a,
\end{equation}
where $a$ is the acceleration of the rope.
\begin{table}[h]
\begin{center}
\begin{tabular}{|c|c|c|c|c|c|}
\hline
System& $M$&$P$&$F$&$u$& $u\,{\rm d}M/{\rm d}t$ \\
\hline \hline
I &$\lambda\,(l\,-\,x)$ &$\lambda\,(l\,-\,x)\,v$& $T(x)$& $v$ &$-\,\lambda\,v^2$\\
\hline
II &$\lambda\,x$ &$\lambda\,x\,v$& $F\,-\,T(x)$&  $v$  &$\,\lambda\,v^2$\\
\hline
I+II &$\lambda\,l$ &$\lambda\,l\,v$& $F$&{---}&{---}\\
\hline
\end{tabular}
\caption{\label{tab:(3)} Summary of  relevant physical quantities for example 3. See  text for notation.}
\end{center}
\end{table}
Regarding  {\bf subsystem II}, we obtain, from (\ref{eq:var})
\begin{equation}\nonumber
\frac{{\rm d}}{{\rm d}t}\,(\lambda\,x\,v)\,=\,-\,T\,(x)\,+\,F\,+\,\lambda \,v^2.
\end{equation}

Using  (\ref{eq:tx3}) in this equation  the acceleration turns out to be $a=F/(\lambda\,l)$. Finally, inserting
this acceleration into (\ref{eq:tx3}), we obtain
\begin{equation}
T\,(x)\,=\,F\,(1\,-\,\frac{x}{l}\,),
\end{equation}
allowing for the correct limits $T\,(x=0)\,=\,F$ and $T\,(x=l)\,=\,0$.

Since  we  have now $u=v$ for both subsystems,  a general equation of type (\ref{eq:ma}) (see also
(\ref{eq:mav1}) and (\ref{eq:mav2})) could be used directly, leading to the correct results.

In this context, as already pointed out by Tiersten \cite{Tiersten} using the concept of flux, equation
(\ref{eq:dp}) could not be used to obtain the tension $T(x)$. In fact, choosing, for instance, the subsystem I we
have: $F\,=\,T\,(x)$, $dP/dt\,=\,-\,\lambda\,v^2\,+\,\lambda\,(l\,-\,x)\,a$, which does not lead to  the correct
expression for $T(x)$.


\section {Conclusions}

In this paper we suggested a method  suitable for  discussing  variable mass
 systems in a straightforward way, and presented three applications.
 Usually  the study of the
systems under consideration is elaborated  over a {\em fixed} number of particles, and so both forms of Newton's
second law, either ${\bf F}\,=\,M\,{\bf a}$ or ${\bf F}\,=\,{\rm d}\,{\bf P}/{\rm d}t$, can be used
\cite{Siegel,Sousa}.

So far as the question of the validity of multi-particle forms of Newton'second law is concerned, three special
cases must be considered:

\begin{itemize}
\item [(i)]  $M\,=\,{\rm constant}$ (in this case ${\bf F}\,=\,{{\rm d}\,{\bf P}}/{{\rm d}\,t}$ and
  ${\bf F}\,=\,M\,{\bf a}$ are equivalent);
 \item [(ii)]  ${\bf u}=0$, i.e. the increment of mass  ${\rm d}\,M$ is at rest;

 and
 \item [(iii)] ${\bf u}\,=\,{\bf v}$.
\end{itemize}

So,  as the general equation (\ref{eq:var})
 shows, ${\bf F}\,=\,{{\rm d}\,{\bf P}}/{{\rm d}\,t}$ is the correct equation of motion  only in two special
 cases (i) and (ii).
For example,  the  raindrop falling through a stationary cloud of droplets \cite{Krane} inserts into the case
(ii). In general, as the present illustrative examples show, conditions (i) and (ii) above are not satisfied and,
consequently,  ${\bf F}\,=\,{{\rm d}\,{\bf P}}/{{\rm d}\,t}$ does not provide the equation of motion of the
corresponding  subsystems I and II.

On the other hand, if the case (iii) applies, a general equation of the type ${\bf F}\,=\,M\,{\bf a}$ can be used
(see (\ref{eq:mav2})).
 Referring, in particular, to example 3, since in this case ${\bf v}\,=\,{\bf u}$ for both
subsystems, they can be solved by such a  general equation.
The same
considerations apply  to subsystem I of the other examples.
  However, let us stress that this does not mean
  that this form of the Newton's second law is more fundamental than   (\ref{eq:dp}) in the context of classical
  mechanics, and that care must be taken when referring to variable mass systems.

\vskip0.3cm
 {\bf Acknowledgments}
\vskip0.3cm

 Work supported by FCT. We would like to thank M. C. Ruivo, L. Brito and A. Blin
for fruitful discussions. We are also grateful to  M. Fiolhais for a careful reading of the manuscript and helpful
suggestions.

\end{document}